\documentclass[preprint,12pt,a4paper]{elsarticle}

\usepackage{amssymb}
\usepackage{amsmath}
\usepackage{graphicx}
\usepackage{array}
\usepackage{booktabs}
\usepackage{listings}
\usepackage{xcolor}
\usepackage{hyperref}
\usepackage[T1]{fontenc}
\usepackage{microtype}
\hypersetup{
    pdftitle={py-ARGOS: A Python package for automatic, statistically principled discovery of ordinary differential equations from data},
    pdfauthor={Yuzheng Zhang, Rui Carvalho}
}

\let\origunderscore\_
\renewcommand{\_}{\origunderscore\allowbreak}

\definecolor{codegreen}{rgb}{0.0,0.5,0.0}
\definecolor{codegray}{rgb}{0.5,0.5,0.5}
\definecolor{codepurple}{rgb}{0.58,0.0,0.82}
\definecolor{codeblue}{rgb}{0.10,0.10,0.65}
\lstdefinestyle{py}{
    language=Python,
    basicstyle=\ttfamily\footnotesize,
    keywordstyle=\color{codeblue},
    commentstyle=\color{codegreen}\itshape,
    stringstyle=\color{codepurple},
    numbers=left,
    numberstyle=\tiny\color{codegray},
    numbersep=7pt,
    breaklines=true,
    breakatwhitespace=false,
    showstringspaces=false,
    aboveskip=1.0em,
    belowskip=1.0em,
    captionpos=b
}
\begin{document}

\begin{frontmatter}

    \title{py-ARGOS: A Python package for automatic, statistically principled discovery of ordinary differential equations from data}

    \author[dur]{Yuzheng Zhang}
    \author[dur]{Rui Carvalho\corref{cor1}}
    \ead{rui.carvalho@durham.ac.uk}
    \address[dur]{Department of Engineering, Durham University, Durham, United Kingdom}
    \cortext[cor1]{Corresponding author.}

    \begin{abstract}
        Recovering the governing equations of a dynamical system from data is a
        central problem in data-driven science, yet the available Python tooling
        forces a compromise among automation, statistical rigor, and reliability.
        We present py-ARGOS, a Python package implementing two automatic
        algorithms, bootstrap ARGOS and Bayesian-ARGOS, behind a uniform
        interface. Both are statistically principled throughout: automated model
        selection reduces the candidate library, each coefficient is reported
        with a bootstrap confidence or Bayesian credible interval that quantifies
        its uncertainty, and a term is retained when this interval excludes zero.
        Built-in diagnostics additionally flag unreliable fits. On chaotic
        benchmark systems, the two algorithms improve on the SINDy family's
        identification performance.
    \end{abstract}


\end{frontmatter}


\section*{Required Metadata}

\section*{Current code version}

\begin{table}[!ht]
    \begin{tabular}{|l|>{\raggedright\arraybackslash}p{4.8cm}|>{\raggedright\arraybackslash}p{6.6cm}|}
        \hline
        \textbf{Nr.} & \textbf{Code metadata description}                               & \textbf{Value} \\
        \hline
        C1           & Current code version                                             & v0.1.0         \\
        \hline
        C2           & Permanent link to code/repository used for this code version     &
        \url{https://github.com/yuzhengv/py-ARGOS}                                                       \\
        \hline
        C3           & Legal Code License                                               &
        MIT                                                                                              \\
        \hline
        C4           & Code versioning system used                                      &
        git                                                                                              \\
        \hline
        C5           & Software code languages, tools, and services used                &
        Python (3.11--3.12); adelie, ArviZ, bambi, NumPy, scikit-learn, SciPy, statsmodels; optional nutpie,
        JAX (GPU)                                                                                        \\
        \hline
        C6           & Compilation requirements, operating environments \& dependencies &
        \texttt{pip install -e .} (editable); optional extras
        \texttt{[accelerator]}, \texttt{[gpu]}, \texttt{[jax]}, \texttt{[dev]};
        Linux/macOS/Windows                                                                              \\
        \hline
        C7           & Link to developer documentation/manual                           &
        \url{https://github.com/yuzhengv/py-ARGOS/blob/main/README.md}                                   \\
        \hline
        C8           & Support email for questions                                      &
        rui.carvalho@durham.ac.uk                                                                        \\
        \hline
    \end{tabular}
    \caption{Code metadata.}
    \label{tab:metadata}
\end{table}

\section{ Motivation and significance}

Governing equations describe how a dynamical system evolves in time, and recovering them directly from measured data is a fundamental inverse problem across physics, biology, engineering, and the climate sciences. This work concerns ordinary differential equations: given observations of a state vector $x(t)\in\mathbb{R}^{m}$, the goal is to infer the right-hand side $f$ of $\dot{x}=f(x)$, or equivalently the component equations $\dot{x}_j = f_j(x)$, $j=1,\dots,m$. Deriving $f$ from first principles is often laborious and demands substantial domain expertise, so library-based sparse regression methods, such as SINDy~\cite{bruntonDiscoveringGoverningEquations2016}, offer a data-driven alternative. These methods first assemble a library of candidate functions, typically polynomials, trigonometric functions, and other nonlinear terms, and assume that each $f_j$ is well represented by only a few of them. Identification then reduces to a sparse regression problem: coefficients are estimated for all candidate terms, and only those terms whose coefficients are non-negligible or statistically significant are retained. The recovered equations are usually parsimonious and mechanistically interpretable, lending themselves to prediction, control, and scientific insight.

Among these methods, sparse identification of nonlinear dynamics (SINDy)~\cite{bruntonDiscoveringGoverningEquations2016} and its variants are the dominant approach, and their reference implementation, PySINDy~\cite{desilvaPySINDyPythonPackage2020,kaptanoglu2022}, is the field's principal actively maintained Python package. PySINDy is broad: beyond the original sequentially-thresholded least squares it bundles the major SINDy extensions, notably AIC-based model selection~\cite{manganModelSelectionDynamical2017}, which lessens the dependence on the manually chosen threshold of SINDy's original selection step, and Ensemble-SINDy~\cite{faselEnsembleSINDyRobustSparse2022}, which adds empirical uncertainty quantification. This combination of breadth and maintenance has made the SINDy family in general, and PySINDy in particular, the most popular and most convenient choice for library-based equation discovery.

Research beyond the SINDy core nonetheless continues to improve on it along several axes: ADAM-SINDy~\cite{vikneshADAMSINDyEfficientOptimization2025} jointly optimizes the nonlinear parameters embedded in library terms together with their coefficients; WyNDA~\cite{hasanWyNDAMethodDiscover2024} identifies governing equations online through an adaptive observer, without iterative optimization; ARGOS~\cite{eganAutomaticallyDiscoveringOrdinary2024} automates the entire discovery pipeline, including denoising, differentiation, and sparse regression, and selects terms through bootstrap confidence intervals rather than a manually tuned threshold, quantifying coefficient uncertainty in the process; and Bayesian-ARGOS~\cite{zhang2026} makes that principled inference efficient, using a fast frequentist screen to reduce the library before confining Bayesian inference to the reduced set, so that posterior credible intervals are obtained at a fraction of the bootstrap's cost. These advances are difficult to reach in practice, however: each is released as paper-accompanying code (MATLAB, R, or unmaintained scripts) rather than as a general, installable Python package, and so stays out of reach for most practitioners.

Beyond this accessibility barrier, many discovery algorithms retain two limitations that affect reproducibility and reliability. First, term selection often depends on a user-specified sparsity threshold or inclusion cutoff, such as a coefficient threshold in sequential thresholding or a cutoff used to interpret ensemble inclusion probabilities. Because these quantities are not uniquely determined by the data, the recovered equation can change with analyst choices. Second, most implementations do not routinely expose built-in diagnostics for assessing whether an apparently sparse model is reliable. Multicollinearity can make the selected support unstable, influential observations can drive individual coefficients, and a misspecified candidate library can force the regression to select compensating terms. Without diagnostics for these failure modes, the software may still return a plausible but incorrect equation without warning.

The ARGOS line of work addresses this gap at the methodological level. ARGOS (Automatic Regression for Governing Equations)~\cite{eganAutomaticallyDiscoveringOrdinary2024} combines Savitzky--Golay denoising and differentiation, cross-validated regularization-path selection for lasso and adaptive-lasso regression, and information-criterion model selection into a single automated screening that reduces the candidate library to a sparse model, then applies a bootstrap-confidence-interval rule that retains a term only when its interval excludes zero, attaching an uncertainty estimate to every selected coefficient and dispensing with any manually chosen threshold. Bayesian-ARGOS~\cite{zhang2026} keeps this automated screening, with minor refinements, but replaces the bootstrap with Bayesian inference on the reduced candidate library: terms are then selected by posterior credible intervals, the probabilistic formulation makes available reliability diagnostics that non-probabilistic pipelines cannot provide, and confining inference to the screened library keeps the procedure roughly two orders of magnitude cheaper than full bootstrap ARGOS. Beyond this statistical rigor, both methods also raise identification success rates over the SINDy family in several scenarios spanning different time-series lengths and noise levels. As software, however, neither is available to Python users through a single, installable package: ARGOS provides only an R codebase, and Bayesian-ARGOS's Python implementation has not yet been packaged.

Here we present py-ARGOS, the companion Python package that closes this software gap, bringing both algorithms together behind a single interface. The user supplies a (generally noisy) state time series $\tilde X\in\mathbb{R}^{n\times m}$ with its sampling step $\Delta t$ and constructs a candidate library, a step that automatically smooths and differentiates the trajectories; calling \texttt{run()} on one of two estimator classes, \texttt{BootstrapArgos} or \texttt{BayesianArgos}, then carries out the frequentist screening and term selection without manual intervention, returning the identified symbolic equations together with simulated trajectories. This design resolves both limitations directly. Retaining a term only when its bootstrap or posterior interval excludes zero makes selection a formal statistical decision at a stated level: it needs no hand-set threshold or inclusion cutoff, and it does not vary with analyst choices. The package also ships the failure diagnostics that competing tools omit, so an unreliable identification need not pass unnoticed: variance inflation factors flag multicollinearity, Pareto-smoothed importance-sampling leave-one-out cross-validation (PSIS-LOO) flags influential observations, and posterior-predictive residual checks flag misspecification, each available on demand so that the model can be revised or the data reprocessed before it is trusted. Being written in Python, the entire workflow composes directly with the existing scientific-Python and probabilistic-programming ecosystem, and can be integrated into larger identification pipelines built around neural networks, whether to reduce dimensionality and recover the dynamics in a latent space~\cite{gaoSparseIdentificationNonlinear2025} or to extend identification to more complex settings such as network dynamics~\cite{gaoLearningInterpretableDynamics2024}.

\section{Software description}

py-ARGOS is a Python package for discovering ordinary differential equations from data. It unifies the two ARGOS-family algorithms, bootstrap ARGOS and Bayesian-ARGOS, within a single estimator interface that couples their statistically principled, threshold-free term selection with built-in failure diagnostics for Bayesian fits. The distribution \texttt{pyargos-equation-discovery} (import \texttt{py\_argos}) is MIT-licensed and supports Python 3.11--3.12. Its dependencies follow the discovery pipeline. NumPy and SciPy handle numerics, filtering, and ODE integration, and scikit-learn builds and scales the polynomial libraries. The frequentist screening then uses \texttt{adelie}, the Python analogue of R's \texttt{glmnet}~\cite{yang_adelie,friedmanRegularizationPathsGeneralized2010}, which fits the cross-validated penalized (lasso/ridge) regressions, with statsmodels providing the OLS refits, BIC selection, and variance-inflation diagnostics. The Bayesian stage builds models with Bambi~\cite{capretto2022} on the PyMC/PyTensor stack~\cite{abrilpla2023}, with ArviZ~\cite{kumar2019} for posterior summaries and diagnostics, optionally accelerated by the external \texttt{nutpie} sampler (numba or JAX backend).

\subsection{Software architecture}

\begin{figure}[!ht]
    \centering
    \includegraphics[width=0.62\textwidth]{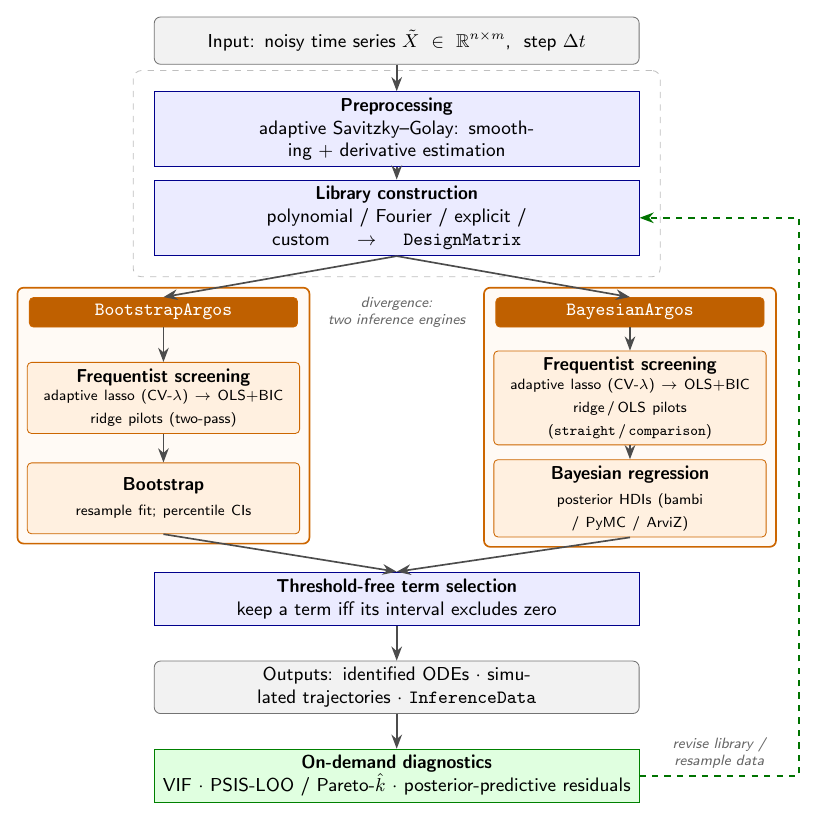}
    \caption{Architecture of py-ARGOS. Preprocessing and library construction form a shared front-end; the two algorithms then run separate inference engines, each opening with a frequentist screening stage (structurally identical but configured differently per algorithm) and closing with a distinct final inference step---bootstrap resampling in \texttt{BootstrapArgos}, Bayesian regression in \texttt{BayesianArgos}---before rejoining at threshold-free, interval-based term selection. Outputs can be inspected with on-demand diagnostics, which can inform refinements to library construction and resampling.}
    \label{fig:arch}
\end{figure}

The two algorithms share a common front-end (preprocessing and library construction) and then run separate inference engines. Each engine begins with a frequentist screening stage, structurally identical but configured differently for each algorithm, and ends with a distinct final inference step (Fig.~\ref{fig:arch}). We describe this pipeline first (Section~\ref{sec:pipeline}) and then the estimator classes that expose its inference engines (Section~\ref{sec:classes}).

\subsubsection{Processing pipeline}\label{sec:pipeline}

The pipeline runs in three stages: preprocessing and library construction, both fully shared, and an inference engine in which the two algorithms diverge. First, measured trajectories are smoothed and differentiated by an adaptive Savitzky--Golay filter~\cite{savitzkySmoothingDifferentiationData1964}, whose window length is chosen automatically by minimizing a reconstruction-error criterion (\texttt{preprocess\_observations}, \texttt{sg\_optimal\_combination}). From the smoothed state, a candidate library is assembled by one of three builders of increasing flexibility: \texttt{build\_design\_matrix} generates polynomial, Fourier, or mixed libraries automatically; \texttt{build\_specific\_design\_matrix} accepts explicit terms through a small, safe expression parser (supporting products such as \texttt{x1x2} and \texttt{x1\^{}2x3} as well as trigonometric and exponential functions); and \texttt{build\_custom\_design\_matrix} admits an arbitrary, pre-evaluated candidate matrix. All three yield a common, normalized design-matrix representation. Each inference engine then opens with a frequentist screening stage that reduces this library through a two-pass adaptive-lasso procedure~\cite{zouAdaptiveLassoIts2006,tibshiraniRegressionShrinkageSelection1996}: each pass pairs a cross-validated adaptive lasso with OLS refits over a fine-grained threshold grid, the best selected by the Bayesian information criterion~\cite{schwarzEstimatingDimensionModel1978}, and between passes the library is trimmed to the highest polynomial degree retained in the first. The two algorithms share the structure of this screening but configure it differently, setting the adaptive penalties from ridge~\cite{hoerlRidgeRegressionBiased1970} or OLS pilot weights. They diverge more substantially at the final inference step: \texttt{BootstrapArgos} resamples the screened fit and retains terms whose bootstrap confidence interval excludes zero, whereas \texttt{BayesianArgos} performs Bayesian regression on the retained terms and keeps those whose posterior credible interval, the highest-density interval (HDI), excludes zero. Both then return the identified symbolic equations and simulated trajectories, and \texttt{BayesianArgos} additionally exposes the full posterior as an ArviZ \texttt{InferenceData} object together with the on-demand failure diagnostics (Section~\ref{sec:diagnostics}). The frequentist screening of bootstrap ARGOS follows the ARGOS reference implementation rather than the algorithm exactly as described in the original paper, incorporating a small, empirically motivated adjustment.

\subsubsection{Estimator classes}\label{sec:classes}

Whereas preprocessing and library construction are provided as standalone functions, the inference engine is exposed through two estimator classes. \texttt{BootstrapArgos} performs bootstrap-confidence-interval identification, applying ridge-pilot screening followed by a bootstrap stage that refits the adaptive lasso on each resampled replicate. \texttt{BayesianArgos} performs Bayesian identification and exposes two screening modes: \texttt{straight} reproduces the Bayesian-ARGOS procedure, using ridge-derived adaptive weights in the first pass and OLS-derived weights in the second, whereas \texttt{comparison} runs the screening twice, once with ridge weights throughout and once with OLS, and keeps the lower-BIC fit. Both modes feed identical Bayesian inference and credible-interval selection, differing only within the frequentist screening. We retain the \texttt{comparison} mode because, although \texttt{straight} was generally more accurate across the systems we tested, \texttt{comparison} outperformed it on certain ones. A companion class, \texttt{BayesianArgosAnalysis}, returns the intermediate screening objects (the adaptive-lasso fits and full regularization paths) for inspection. A shared simulation mixin serves both estimators, whereas the failure diagnostics are exposed through a Bayesian-facing diagnostics mixin.

\subsection{Software functionalities}

Building on this architecture, the functionality the package exposes to the user is deliberately compact: a uniform identification interface (Section~\ref{sec:identification}), configurable bootstrap and Bayesian inference with parallel computation (Section~\ref{sec:computation}), utilities for simulating and summarizing the identified model (Section~\ref{sec:simulation}), and built-in failure diagnostics (Section~\ref{sec:diagnostics}).

\subsubsection{Equation identification}\label{sec:identification}

The two estimator classes share a constructor-plus-\texttt{run()} interface and identify either the whole system or individual equations (\texttt{run}, \texttt{run\_each\_equation}). Because terms are retained on the basis of their bootstrap or posterior intervals, identification does not depend on an analyst-chosen sparsity threshold, and the only quantities the user ordinarily sets are the interval level (\texttt{ci\_level}), the number of bootstrap samples (\texttt{num\_samples}), and the MCMC settings. The candidate library may be polynomial, trigonometric, mixed, explicitly enumerated, or fully user-defined, so the same interface accommodates both routine polynomial discovery and problems that require bespoke nonlinear terms.

\subsubsection{Inference and parallel computation}\label{sec:computation}

The two engines expose different inference controls: bootstrap ARGOS needs little beyond the number of resampling replicates, whereas Bayesian-ARGOS adds priors and sampler settings. These priors are by default weakly-informative, automatically scaled in the style of rstanarm~\cite{gelmanWeaklyInformativeDefault2008}, namely $\beta_k\sim\mathcal{N}(0,(2.5\,s_y/s_{x_k})^2)$, intercept $\mathcal{N}(m_y,(2.5\,s_y)^2)$, and $\sigma\sim\mathrm{Exp}(1/s_y)$, which the user may override. Computation is parallelized within each engine: bootstrap ARGOS distributes its resampling replicates across cores, while Bayesian-ARGOS runs its MCMC chains across cores and can additionally call the optional \texttt{nutpie} sampler with \texttt{numba} (CPU) or JAX (GPU) backends. The GPU backend helps only when the per-step gradient evaluations of the log-posterior are compute-bound (large $n$, large libraries), since the NUTS tree-building control flow itself remains on the CPU. The larger, algorithmic speedup of Bayesian-ARGOS over bootstrap ARGOS is quantified in Section~\ref{sec:impact}.

\subsubsection{Simulation and results extraction}\label{sec:simulation}

The \texttt{simulate()} method integrates the identified system with SciPy, and a family of result-extraction methods (\texttt{get\_identified\_model\_from\_all\_equations}, \texttt{get\_frequentist\_results}, and \texttt{expressions\_for\_simulation}) return tidy coefficient tables and ready-to-simulate symbolic expressions, while each fit's results object exposes a \texttt{results\_summary} of posterior or bootstrap confidence intervals; three-dimensional trajectory plotting is also provided.

\subsubsection{Failure diagnostics}\label{sec:diagnostics}

A distinguishing capability is a suite of statistical diagnostics that target the specific ways a sparse identification can fail, computed on demand after a fit rather than as part of it. Variance inflation factors (\texttt{compute\_vif}/\texttt{plot\_vif}) flag multicollinearity above the conventional threshold of ten~\cite{marquardtRidgeRegressionPractice1975,obrienCautionRegardingRules2007a}; PSIS-LOO Pareto-$\hat k$ values (\texttt{compute\_pareto\_k}/\texttt{plot\_pareto\_k}) flag influential observations above $0.7$~\cite{vehtariPracticalBayesianModel2017}; and posterior-predictive residual checks (\texttt{compute\_residuals}/\texttt{plot\_residuals}) reveal model misspecification. Because the Pareto-$\hat k$ and residual diagnostics rest on the posterior, the suite accompanies the Bayesian estimator, whereas the variance-inflation check depends only on the design matrix and applies to any candidate library. Each diagnostic separates a compute step, which returns the underlying values, from a plotting step, so results can be used programmatically or inspected visually. Together they give the user a principled basis for judging whether an identified model can be trusted, a capability competing packages rarely expose.

\subsection{Limitations}

Against these capabilities, several limitations remain, both methodological and practical. Methodologically, py-ARGOS inherits the core assumptions of regression-based identification, in particular that the true terms must be present in the candidate library. In practice, performance depends on data quantity and signal quality, sometimes counterintuitively: near-zero noise can violate homoscedasticity, and dense sampling on a chaotic attractor can inflate multicollinearity, though both are diagnosable with the built-in tools. Computationally, bootstrap ARGOS becomes expensive at large $n$, which motivates the cheaper Bayesian-ARGOS. On the software side, \texttt{adelie} currently requires \texttt{numpy<2} and, on ARM64 Linux, must be built from source.

\section{Illustrative examples}\label{sec:examples}

The two examples below run py-ARGOS end to end on chaotic systems with known governing equations, one for each estimator class: Bayesian-ARGOS with an automatically generated polynomial library, together with its diagnostics, on the R\"ossler system; and bootstrap ARGOS with an explicit, user-specified library on the Thomas system.

\subsection{Bayesian identification and diagnostics}

\begin{figure}[!ht]
    \centering
    \includegraphics[width=0.92\textwidth]{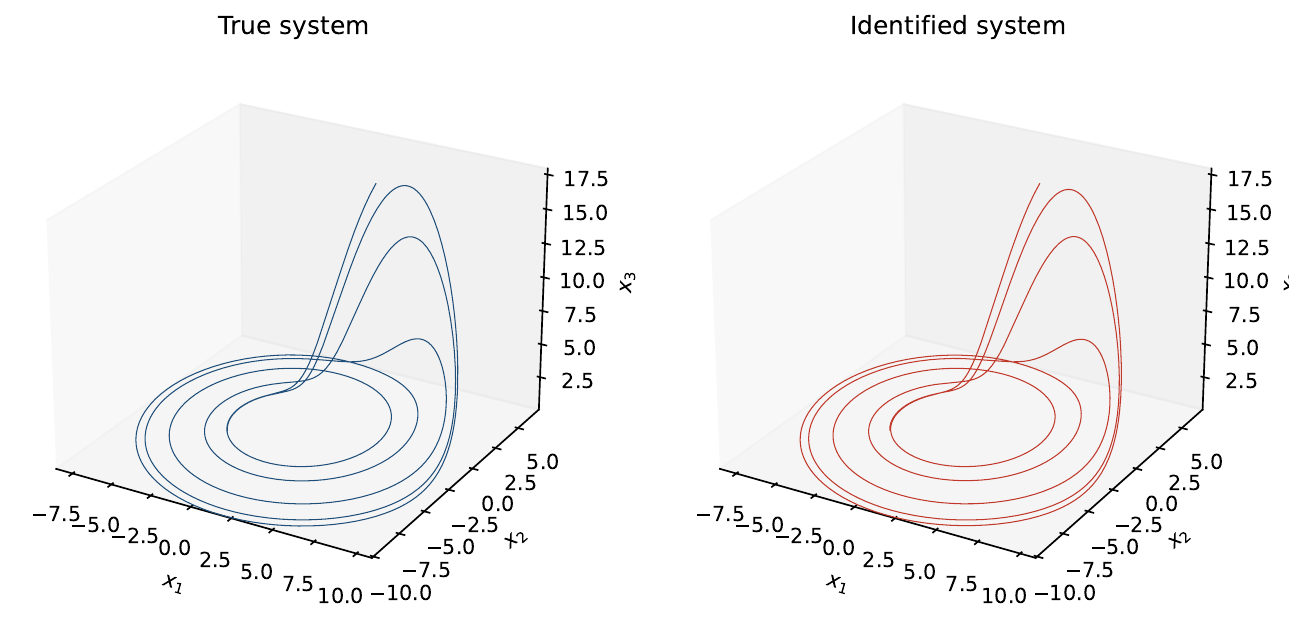}
    \caption{Validation by trajectory recovery. The identified system (right), simulated from the equations recovered by py-ARGOS, closely matches the true R\"ossler attractor (left). From data at SNR $\approx 49$~dB the recovered model is $\dot x_1=-1.00\,x_2-1.00\,x_3$, $\dot x_2=1.00\,x_1+0.20\,x_2$, $\dot x_3=0.20+1.00\,x_1x_3-5.70\,x_3$: every estimated nonzero coefficient within $0.2\%$ of truth, with no spurious terms.}
    \label{fig:traj}
\end{figure}

We discover the chaotic R\"ossler system
\begin{equation}
    \begin{aligned}
        \dot x_1 & = -x_2-x_3,             \\
        \dot x_2 & = x_1+0.2\,x_2,         \\
        \dot x_3 & = 0.2+x_3\,(x_1 - 5.7),
    \end{aligned}
\end{equation}
from noisy data. Listing~\ref{lst:rossler} generates a trajectory at SNR $\approx 49$~dB, builds a polynomial library up to degree five, and fits \texttt{BayesianArgos} in \texttt{"straight"} mode using 90\% highest-density intervals (HDIs) for term selection. It then returns the identified equations and simulates the recovered system. Figure~\ref{fig:traj} places the identified attractor beside the true one; their agreement validates that the recovered equations reproduce the observed dynamics.

\begin{lstlisting}[caption={Discovering the R\"ossler system with Bayesian-ARGOS.},label={lst:rossler}]
from py_argos import BayesianArgos
from py_argos import design_matrix as agu
from py_argos import simulation as ags

# 1. Generate a noisy Rossler trajectory (true system shown for reference).
init = ags.generate_initial_value_df(seed=100, num_init_samples=100,
        x1_range=(-10, 10), x2_range=(-10, 10), x3_range=(0, 20), num_columns=3)
x_t = ags.generate_noisy_dynamical_systems(
        variable_coeff=[[-1, -1], [1, 0.2], [0.2, 1, -5.7]],
        variable_names=[["x2", "x3"], ["x1", "x2"], ["", "x1x3", "x3"]],
        n=10**3.5, dt=0.01, init_conditions=init.iloc[95].values, snr=49)

# 2. Build a degree-5 polynomial library (adaptive SG smoothing + derivatives).
dm = agu.build_design_matrix(x_t=x_t, dt=0.01, library_degree=5,
        sg_poly_order=4, library_type="poly")

# 3. Fit Bayesian-ARGOS; terms are kept iff their 90% HDI excludes zero.
model = BayesianArgos(design_matrix=dm, custom_prior=True, accelerator=False)
model.run(mode="straight", parallel="yes", ncpus=None, ci_level=0.90)
model.get_identified_model_from_all_equations()   # tidy coefficient table

# 4. Simulate the identified system and compare trajectories.
x_sim = model.simulate(n=10**3.5, dt=0.01, init_conditions=init.iloc[95].values)
ags.plot_trajectory_3d(x_sim, "Rossler_identified")
\end{lstlisting}

\begin{figure}[!ht]
    \centering
    \includegraphics[width=\textwidth]{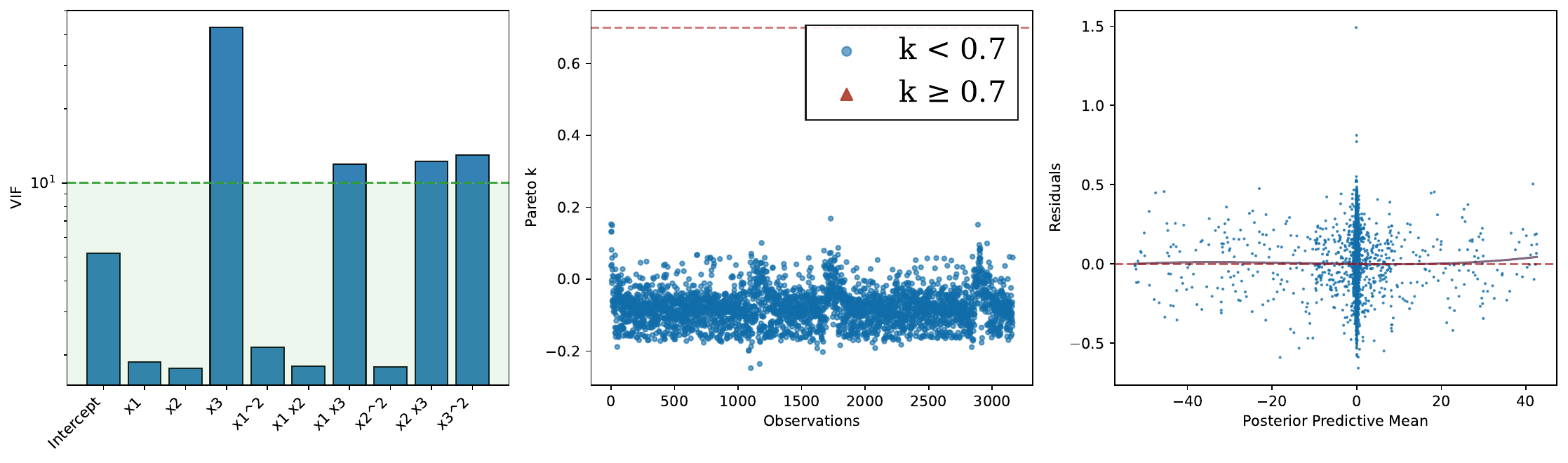}
    \caption{Diagnostics for the R\"ossler fit of Figure~\ref{fig:traj} (equation $\dot x_3$). Left: variance inflation factors on a log scale; several library terms lie above the threshold-10 line (dashed), with the $x_3$ term at $\approx 43$ and the $x_1x_3$, $x_2x_3$, and $x_3^2$ terms near 10--13, indicating strong multicollinearity in the polynomial library. Middle: Pareto-$\hat k$ for all observations, every value below the 0.7 line (no influential observations). Right: posterior-predictive residuals vs.\ mean, scattered about zero with no systematic structure.}
    \label{fig:diag}
\end{figure}

Listing~\ref{lst:diag} runs the three diagnostics on the same R\"ossler fit. We apply them deliberately to a successful identification, to show that the suite characterizes a fit rather than merely confirms its success. Figure~\ref{fig:diag} summarizes the outcome. The variance inflation factors reveal the multicollinearity expected for a polynomial library sampled along a chaotic attractor: several terms exceed the conventional threshold of 10, and the factor for $x_3$ reaches approximately 43, reflecting its near-collinearity with the retained $x_1x_3$, $x_2x_3$, and $x_3^2$ terms. The other two diagnostics are clean: all Pareto-$\hat k$ values fall below 0.7, indicating no influential observations, and the posterior-predictive residuals show no systematic structure. Read together, the panels identify a collinear library as the fit's only genuine weakness and show that neither influential observations nor model misspecification is present. With this relatively large, low-noise sample, the collinearity is benign, and the correct system is still recovered. More generally, in more challenging regimes the same diagnostics would attribute an unsuccessful identification to a specific cause, whether collinearity, an influential observation, or a misspecified library, so that the reason for failure is diagnosed rather than left opaque~\cite{zhang2026}.

\begin{lstlisting}[caption={On-demand failure diagnostics for a fitted model.},label={lst:diag}]
# After model.run(...): compute returns data; plot draws one panel each.
vif      = model.compute_vif(equation_num=3)        # multicollinearity
pareto_k = model.compute_pareto_k(equation_num=3)   # influential points (PSIS-LOO)
resid    = model.compute_residuals(equation_num=3)  # posterior-predictive residuals

import matplotlib.pyplot as plt
fig, ax = plt.subplots(1, 3, figsize=(18, 5))
model.plot_vif(equation_num=3, ax=ax[0])
model.plot_pareto_k(equation_num=3, ax=ax[1])
model.plot_residuals(equation_num=3, ax=ax[2])
\end{lstlisting}

\subsection{Bootstrap confidence interval selection}

\begin{figure}[!ht]
    \centering
    \includegraphics[width=\textwidth]{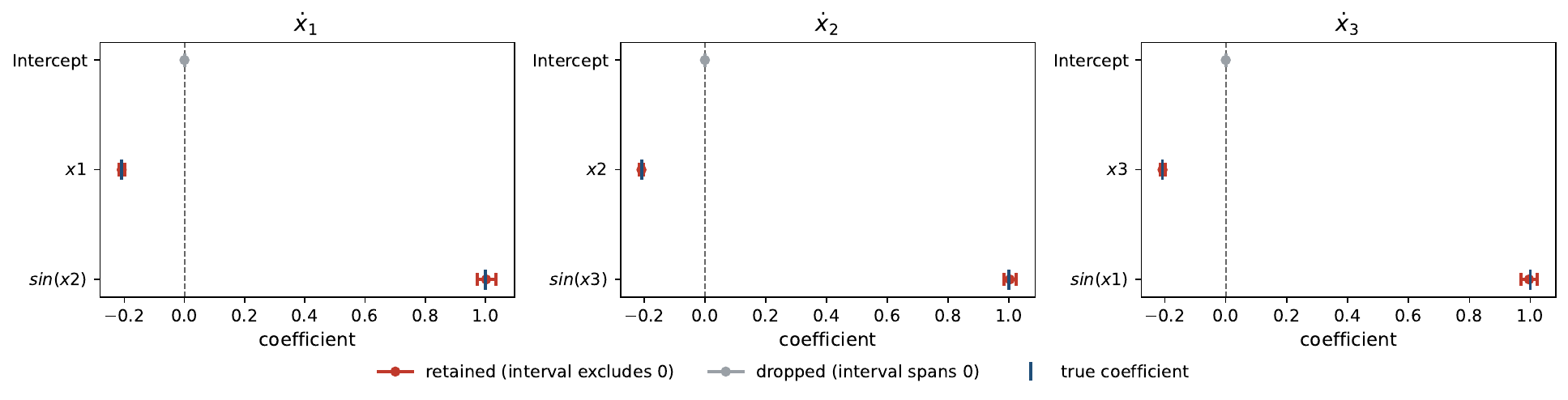}
    \caption{Bootstrap interval-based selection on the Thomas system (\texttt{BootstrapArgos}, explicit user-specified library, SNR $\approx 35$~dB, $B=2000$). Each panel shows, for one equation, the bootstrap coefficient estimate (dot) and 90\% confidence interval (bar) of every screened candidate term, including the automatically added intercept; a term is retained iff its interval excludes zero (dashed line). In each equation the two true terms---the sine coupling and the linear damping---are retained with intervals bracketing the true coefficients (blue ticks), whereas the intercept is correctly dropped.}
    \label{fig:thomas}
\end{figure}

We next identify the chaotic Thomas system
\begin{equation}
    \begin{aligned}
        \dot x_1 & = \sin x_2-b\,x_1, \\
        \dot x_2 & = \sin x_3-b\,x_2, \\
        \dot x_3 & = \sin x_1-b\,x_3,
    \end{aligned}
\end{equation}
with $b=0.208186$. Using \texttt{build\_specific\_design\_matrix}, Listing~\ref{lst:thomas} constructs a user-specified library of six nonconstant terms, the three state variables and their sines; an intercept is added automatically and need not be specified. The estimator \texttt{BootstrapArgos} is then run with $B=2000$ resamples at a 90\% confidence level. At SNR $\approx 35$~dB, it recovers the two true nonconstant terms in each equation, estimates their coefficients to within $0.5\%$ of the true values, and excludes the intercept. Figure~\ref{fig:thomas} shows the resulting per-term confidence intervals, from which selection follows directly: a term is retained only if its interval excludes zero, with no threshold to tune.

\begin{lstlisting}[caption={Discovering the trigonometric Thomas system with bootstrap ARGOS.},label={lst:thomas}]
from py_argos import BootstrapArgos
from py_argos import design_matrix as agu
from py_argos import simulation as ags

# 1. Generate a noisy Thomas trajectory (b = 0.208186), SNR ~ 35 dB.
init = ags.generate_initial_value_df(seed=100, num_init_samples=100,
        x1_range=(-1, 1), x2_range=(-1, 1), x3_range=(-1, 1), num_columns=3)
x_t = ags.generate_noisy_dynamical_systems(
        variable_coeff=[[1, -0.208186], [1, -0.208186], [1, -0.208186]],
        variable_names=[["sin(x2)", "x1"], ["sin(x3)", "x2"], ["sin(x1)", "x3"]],
        n=10**3.5, dt=0.01, init_conditions=init.iloc[95].values, snr=35)

# 2. Explicit candidate library: the state variables and their sines.
dm = agu.build_specific_design_matrix(x_t=x_t,
        specific_terms=["x1", "x2", "x3", "sin(x1)", "sin(x2)", "sin(x3)"],
        dt=0.01, sg_poly_order=4)

# 3. Bootstrap ARGOS: keep a term iff its 90% confidence interval excludes 0.
model = BootstrapArgos(design_matrix=dm, num_samples=2000, random_state=0)
model.run(mode="straight", parallel="yes", ncpus=None, ci_level=0.90)
model.get_identified_model_from_all_equations()   # tidy coefficient table
\end{lstlisting}

These two examples are representative rather than exhaustive: the package's \texttt{examples/} directory provides runnable scripts that demonstrate both algorithms across a variety of diagnostic cases and library types.

\section{Impact}\label{sec:impact}

py-ARGOS provides statistically principled equation discovery within the Python ecosystem. For each candidate term, it estimates the associated coefficient together with a bootstrap confidence interval or a Bayesian credible interval, and retains the term only when this interval excludes zero. Term selection thus follows an explicit inferential criterion, rather than an empirical inclusion frequency or a comparison of coefficient magnitude against a separately tuned sparsity cutoff. This formulation allows the evidence for individual terms to be assessed and discovered models to be reported with quantified parameter uncertainty.

Three on-demand diagnostics extend this rigor from term selection to model reliability. Rather than accepting a fitted model without further scrutiny, the user can inspect each diagnostic and address the issue it identifies: examining influential observations flagged by the Pareto-$\hat k$ statistic, pruning collinear candidate terms indicated by high variance inflation factors, or revising a candidate library that posterior-predictive residual checks indicate is misspecified. py-ARGOS thereby supports not only equation discovery but also assessment of model adequacy~\cite{zhang2026}, so that a discovered equation can be justified when it underlies a scientific claim or a downstream control application.

Implementing these methods in Python also has practical benefits. py-ARGOS makes the ARGOS and Bayesian-ARGOS pipelines, previously available only in R or as an unpublished implementation, accessible within an ecosystem in which they interoperate with established scientific-Python and probabilistic-programming tools. Once an algorithm and a candidate library have been specified, the regularization path, library trimming, and term selection proceed automatically, and results are returned as ArviZ \texttt{InferenceData}, so that the full posterior remains available for subsequent uncertainty propagation, model comparison, or visualization. Bayesian-ARGOS additionally keeps uncertainty-quantified discovery tractable at the large sample sizes at which bootstrap resampling becomes computationally costly.

Table~\ref{tab:compare} compares py-ARGOS with the established Python implementation PySINDy and with the original R implementation of ARGOS. py-ARGOS is distinguished by the statistically principled form of its term selection, together with full Bayesian posterior inference and built-in failure diagnostics.

\begin{table}[!ht]
    \centering
    \resizebox{\textwidth}{!}{%
        \begin{tabular}{lccc}
            \toprule
            \textbf{Capability}                                  & \textbf{py-ARGOS} & \textbf{PySINDy}    & \textbf{ARGOS (R)} \\
            \midrule
            Automatic, threshold-free term selection             & Yes               & Partial (AIC-based) & Yes                \\
            Per-term confidence / credible intervals             & Yes               & Partial (ensemble)  & Yes (bootstrap)    \\
            Full Bayesian posterior inference                    & Yes               & No                  & No                 \\
            Built-in failure diagnostics (VIF, PSIS-LOO, resid.) & Yes               & No                  & No                 \\
            Optional GPU acceleration                            & Yes               & No                  & No                 \\
            Language                                             & Python            & Python              & R                  \\
            \bottomrule
        \end{tabular}%
    }
    \caption{Feature comparison of py-ARGOS with the main alternatives.}
    \label{tab:compare}
\end{table}

The quantitative advantages of the underlying methods were established on chaotic benchmarks using the original R implementations, which we cite rather than reproduce: ARGOS outperforms SINDy-with-AIC on three-dimensional systems~\cite{eganAutomaticallyDiscoveringOrdinary2024}, and Bayesian-ARGOS is more data-efficient and more noise-robust than SINDy across most of seven chaotic systems, with identification performance comparable to that of ARGOS but attained at roughly hundredfold lower runtime~\cite{zhang2026}. The utility of Bayesian-ARGOS beyond classical benchmark systems is further illustrated by its integration with SINDy-SHRED~\cite{gaoSparseIdentificationNonlinear2025}, where it identifies the latent dynamics for sea-surface-temperature modeling and, compared with SINDy, raises the valid-identification rate from 60\% to 77\% while reducing the reconstruction error~\cite{zhang2026}.

\section{Conclusions}

py-ARGOS is the first Python package to implement both bootstrap ARGOS and Bayesian-ARGOS, exposing the two algorithms through a uniform interface for automatic, statistically principled discovery of ordinary differential equations. It addresses a specific gap in the otherwise SINDy-dominated Python ecosystem, in which automatic operation, statistically principled uncertainty quantification, and built-in reliability diagnostics had not previously been available together. We have demonstrated the package on chaotic benchmark systems, and its underlying methods extend to real-world applications such as reduced-order modeling of climate data. Planned extensions include incorporating ARGOS-RAL~\cite{liAutomatingDiscoveryPartial2024} to move from ODE to PDE discovery, together with support for control inputs and additional libraries and priors.

\section*{Acknowledgements}


\bibliographystyle{elsarticle-num}
\bibliography{references1,references2,refs-extra}

\end{document}